\documentclass[a4paper]{spie}  

\usepackage[]{graphicx}

\title{Magnetic flux Noise in MgB$_2$ thin films} 

\author{\"Orjan Festin\supit{a}, Peter Svedlindh\supit{a},W.N. Kang\supit{b}, Eun-Mi Choi\supit{c} and Sung-Ik Lee\supit{c}
\skiplinehalf
\supit{a}Department of Materials Science, Uppsala University, P.O. Box 534, SE-751 21 Uppsala, Sweden \\
\supit{b}Department of Physics, Pukung National University, Pusan 608--737, Korea\\
\supit{c}NCRICS, Department of Physics, Pohang University of Science and Technology, Pohang 790--784, Korea
}

\authorinfo{Further author information: (Send correspondence to \"Orjan Festin): E-mail: orjan.festin@angstrom.uu.se, Telephone: +46 (0)18 4713136}

\begin{document} 
  \maketitle 

\begin{abstract}
We have performed flux noise and AC-susceptibility measurements on two $400$ nm thick MgB$_2$ films. Both measurement techniques give information about the vortex dynamics in the sample, and hence the superconducting transition, and can be linked to each other through the fluctuation-dissipation-theorem. The transition widths for the two films are 0.3 and 0.8 K, respectively, and the transitions show a multi step-like behavior in the AC-susceptibility measurements. The same phenomenon is observed in the flux noise measurements through a change in the frequency dependence of the spectral density at each step in the transition. The results are discussed and interpreted in terms of vortices carrying an arbitrary fraction of a flux quantum as well as in terms of different macroscopic regions in the films having slightly different compositions, and hence, different critical temperatures. 
\end{abstract}



\section{INTRODUCTION}
\label{sect:intro}  

Since the discovery of superconductivity in MgB$_2$\cite{Nagamatsu1}, numerous investigations relating to the superconducting and structural properties\cite{Zhu,Serguis} of this material have been performed. The hexagonal lattice of MgB$_2$ consists of two parallel systems of flat layers. One layer contains B atoms in a honeycomb lattice, being mainly responsible for the superconducting properties, the other Mg atoms in a triangular lattice halfway between the B layers. The quasi-two-dimensional (2D) boron planes yield a 2D behavior of the material close to the superconducting transition temperature ($T_{c0}$), as evidenced by resistivity measurements\cite{Sidorenko}, yielding a temperature dependence of the excess conductivity that agrees with predictions of the Aslamazov--Larkin theory\cite{AZL} of fluctuations in 2D superconductors. 

Analysis of thin film samples prepared by different techniques, show that secondary phases such as MgO, MgB$_4$ and MgAl$_2$O$_4$ easily form\cite{Tian}, resulting in  a lowering of $T_{c0}$ as well as a broadening of the superconducting transition. Investigations regarding $T_{c0}$ as a function of Mg content show that the transition temperature decreases with decreasing Mg content\cite{sharma1}. 

First principles calculations\cite{Liu} show that the Fermi surface of MgB$_2$ consists of 2D cylindrical sheets arising from $\sigma$ antibonding states of B $p_{xy}$ orbitals, and 3D tubular networks arising from $\pi$ bonding and antibonding states of B $p_{z}$ orbitals. In this theoretical framework two different energy gaps exist, the smaller connected to the 3D bands and the larger one associated with the superconducting 2D bands. Point contact spectroscopy measurements\cite{Szabo}, performed on clean enough samples, i.e. samples with enough long range microscopic order and few macroscopic defects, have confirmed the theoretical prediction of two band superconductivity. Later, specific heat\cite{Yang,Junod1} and penetration depth measurements\cite{Lemberger} gave further evidence in favour of two band superconductivity. More recently, Babaev\cite{Babaev1} suggested, as a consequence of the two band gap nature, that several vortex configurations can exist in MgB$_2$, such as ordinary vortices with $n\Phi_0$ magnetic flux, but also neutral vortices and vortices carrying an arbitrary fraction of a flux quantum. This theoretical prediction has not yet, however, been verified experimentally.  

In this paper we report on magnetic flux noise and AC-susceptibilty studies on two MgB$_2$ films, performed close to the mean field transition temperature and in the zero magnetic field limit. This implies that properties of spontaneously created vortices have been investigated. The results obtained are compared with theoretical models of 2D superconductors, in which the spontaneously created vortices appear as pairs undergoing a decoupling transition at the Kosterlitz-Thouless transition  temperature, $T_{KT}$\cite{KTB}. 

\section{Theory}
Vortices in a 2D superconductor have, provided that their separation distance is smaller than the effective penetration depth, $\Lambda = \lambda^{2}/t$, where $\lambda$ is the London penetration depth and $t$ the film thickness, a long range interaction depending logarithmically on distance\cite{Minnhagen1,Pearl}. Because of this long range interaction, it is expected that a 2D superconductor should exhibit features reminiscent of a system undergoing a Kosterlitz-Thouless transition.  Below $T_{KT}$ the vortices are "frozen" and bound in pairs consisting of vortices of opposite vorticity. At $T \geq T_{KT}$, vortices decouple provided that their  separation distance is larger than a temperature dependent length scale corresponding to the correlation length $\xi_+$, which in 2D increases exponentially with temperature as $T\rightarrow T_{KT}$, i.e. $\xi_+ \propto \exp[b/\sqrt{T/T_{KT}-1}]$. The correlation length can be linked to a temperature dependent characteristic frequency, $\omega_0$, via $\omega_0 \propto \xi^{-z}$, which gives 
\begin{equation}
	\label{eq:2D}
		\omega_0 \propto exp[-bz/\sqrt{T/T_{KT}-1}].
\end{equation}
If $\omega<\omega_0$, corresponding to a length scale $l > \xi_+$, the dynamics of  vortices will appear uncorrelated, whereas for $\omega>\omega_0$, corresponding to a length scale $l < \xi_+$, the vortex motion will appear correlated\cite{Minnhagen2}. To probe properties of the vortices with respect to temperature and frequency, AC-susceptibility($\chi(\omega,T)$) and flux noise($S_{\Phi}(\omega,T)$) measurements can be used\cite{Rogers,Festin1}. For temperatures close to but above $T_{KT}$, the sample fluctuations/response will be dominated by vortex pairs, whose density decreases as the temperature is increased and more and more vortices unbind. Close to $T_{c0}$ the 2D vortex fluctuations is of Drude type, i.e. the vortices can be described as "free" vortices, yielding for the flux noise spectrum
\begin{equation}
    \label{eq:Drude}
    S_{\Phi}(\omega)\propto \frac{\omega_0}{\omega^{2}+\omega^{2}_{0}}.
\end{equation}
In the intermediate frequency regime, where  $\omega \approx \omega_0$, vortex pairs with a separation distance close to the correlation length dominate the response. The dynamics of these vortices is well parametrized by the Minnhagen phenomenology (MP)  form\cite{Minnhagen1}, describing moving vortex pairs being polarized by a background consisting of other vortex pairs as well as of free vortices, yielding for the flux noise spectrum\cite{Minnhagen2} 
\begin{equation}
	\label{eq:MP}
	S_{\Phi}(\omega)\propto \frac{\omega_0\ln(\omega/\omega_0)}{\omega^2-\omega^{2}_{0}}.
\end{equation}
This description, which has been theoretically verified by calculating flux noise spectra around the KT transition in simulations of the 2D resistively shunted junction model\cite{Minnhagen3}, implies that $S_{\Phi}(\omega)$ is frequency independent for $\omega<\omega_0$, has a crossover to an $\omega^{-3/2}$-behavior in an intermediate frequency regime above $\omega_0$ (vortex pairs) and for high enough frequencies an $\omega^{-2}$ behavior should be observed (free vortices). This equation is valid when the distance between the sample and the pick-up coil is larger than the microscopic length scales on which vortex movements take place in the sample. In such a case, vortices both inside and outside of the pick-up coil contribute to the measured noise, and there is a gradual change in magnetic flux detected by the pick-up coil as a vortex crosses the boundary defined by the coil area.

The flux noise spectrum can be linked to the AC-susceptibility, or the complex conductance, via the fluctuation-dissipation-theorem as\cite{Minnhagen3}
\begin{equation}
	\label{eq:FD}
	\omega S_{\Phi}(\omega,T)\propto T{\textnormal{Im}[\chi(\omega,T)]}.
\end{equation}
For this equation to be valid, it is important that the field applied in the AC-susceptibility measurements is low enough to ensure that no new vortices are created by the field, i.e. that the field is within the linear response regime.

\section{Experimental}
The two samples used in this work were two squared shaped (5 $\times$ 5mm$^2$) $400$ nm thick MgB$_2$ films, grown on (1102)Al$_2$O$3$ substrates\cite{Kang1}. First an amorphous boron film was deposited using a pulsed laser deposition technique. This  film was put in a Nb-tube together with high purity Mg-metal and the tube was then sealed in an Ar atmosphere. Postannealing was performed for $30$ min at $900$  $^{\circ}$C, yielding the best-quality MgB$_2$ films\cite{Kang2}. More detailed information regarding the fabrication technique can be found elsewhere\cite{Kang1,Kang2}. $\theta-2\theta$ and $\phi$ scans show that the films are highly $c$-axis oriented, while the $ab$-directions are randomly distributed in the films. Atomic force microscopy shows that both films have a surface roughness of $\approx$3 nm. The grain diameter in the film is of the order a few tenth of a micrometer. Film A has $T_{c0}=38.9$ K and $\Delta T_{c0} \approx 0.3$ K. Film B was investigated 4 months after film A, implying that this sample is somewhat degraded, with  $T_{c0}=38.1$K and $\Delta T_{c0} \approx 1$ K.

The experimental setup is a Superconducting QUantum Interference Device (SQUID) based setup, designed for complex impedance and flux noise measurements\cite{Magnusson}. The sample space is magnetically shielded, with a $\mu$-metal can and a niobium can, having the background field reduced to $\approx$ $0.1$ mOe. The pick-up coil is a first order gradiometer with a diameter of $1.2$ mm. The AC-drive coil is placed outside the pick-up coil, having a diameter of $2.5$ mm. The coils were placed $\approx 0.5$ mm above the film surface. This implies that the distance between the sample and the coils can be considered as large, therefore making the MP-description\cite{Minnhagen2,Minnhagen3} of the flux noise valid. The magnetic field used in AC-susceptibility measurements was kept at $H_{AC}\approx 0.2$ mOe, which is low enough to ensure a linear response from the samples. The temperature dependence of the AC-susceptibility was studied in the frequency range from $17$ mHz to $51$ kHz. The frequency dependence of the flux noise spectrum was measured in the range from $10$ mHz to $400$ Hz. The construction of the experimental setup allows a very precise temperature control, with a long time temperature stability of $\approx 0.5$ mK.

\section{Results and discussion}

Figure 1 shows for both films the temperature dependence of the real and imaginary parts of the AC-susceptibility for three different frequencies ($17$ mHz, $1.7$ Hz and $170$  Hz). The results of both films indicate that the superconducting transition occurs in  multiple steps. In $\chi'(T)$, the real part of the AC-susceptibility (Figs. 1 (a) and (c)), it is possible to resolve, for both samples, at least four steps in the transition region. Moreover, it can be seen that the transition for the B film is more broad in temperature. The multi-step transition is in $\chi''(T)$ (Figs. 1 (b) and (d)), the dissipative part of the AC-susceptibility, revealed by several peaks, where each peak corresponds to a step in $\chi'(T)$. It was shown in Ref. (19) that the maximum in $\chi''(T)$, for each frequency of the AC-field, gives a reliable estimate of the characteristic frequency $f_{0}(T) = \omega_{0}(T)/2\pi$. This indicates that both MgB$_{2}$ films exhibit multiple characteristic frequencies, since several peaks are observed in each $\chi''(T)$ curve. It is also worth pointing out that even though the width in temperature of the superconducting transition for the here  investigated films differ by a factor of $\approx 3$, the width in temperature of each individual peak, or each step in the transition region, is approximately the same for both films. The temperature decrease of $f_{0}(T)$, as determined by the maxima in $\chi''(T)$ is, however, found to be more rapid for film A comparing to the corresponding decrease observed for film B. 

\begin{figure}
   \begin{center}
   \begin{tabular}{c}
   \includegraphics[height=10cm]{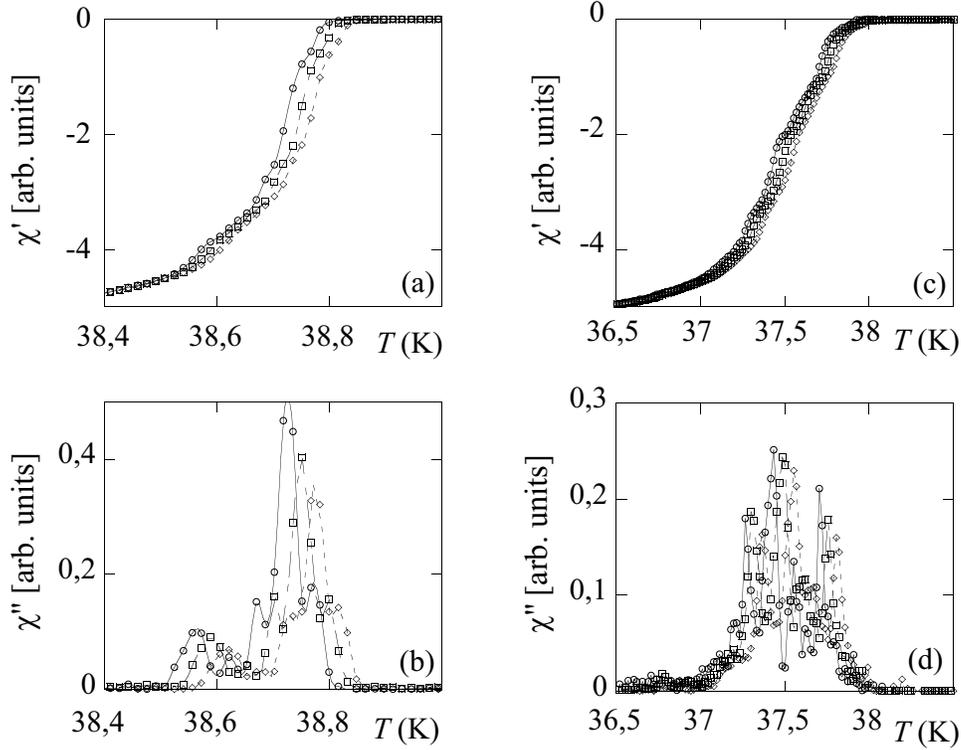}
   \end{tabular}
   \end{center}
   \caption[Figure 1.] 
   { \label{fig:fig1} 
AC-susceptibility vs. temperature for the two MgB$_2$ films. The figure shows AC-susceptibility data for $17$ mHz(open circles), $1.7$ Hz (filled circles) and $170$ Hz (filled squares). The AC-field used in the measurement was $0.2$ mOe.} 
\end{figure}

The only plausible explanation for the multi-step characteristics of the superconducting transition is that there exists different macroscopic regions in each film exhibiting small differences with respect to composition\cite{Festin2}, causing different transition temperatures as well as a widening of the superconducting transition. It is also possible that grain boundaries in the film degrade with time, causing poor superconducting connections between grains, and in the end a non-percolative superconducting system. Film A was investigated freshly prepared, while film B was investigated 4 months after preparation, being stored partly in air and partly in a desiccator. The broadening of the transition in film B is a clear indication of film degradation. Moreover, film A was remeasured 6 months after preparation, and this time the film showed a comparably broad superconducting transition with a transition width of  $\Delta T_{c0} \approx$ 2.5 K.

Theoretical investigations suggest that fractional vortices should occur in MgB$_2$ as a consequence of two band superconductivity, provided that the interband coupling is small \cite{Babaev1,Babaev2}. If one writes the expression for the individual supercurrents, distinguished by index $i =1,2$, in the two-band model as ${\bf j}_i=\frac{i e}{m_i}|\Psi_{i}|^2\nabla \phi_i -\frac{e^2}{4 m_i}|\Psi_i|^2{\bf A}$
it follows that even if there are gradients of the phase variable $\phi_1$, while there are no gradients of $\phi_2$, coupling by $\bf A$ induces a current in the condensate $\Psi_2$ flowing in opposite direction to the current of $\Psi_1$. For a  vortex where $\Delta \phi_1= 2 \pi$, such a current partially compensates the magnetic flux induced by $\Psi_1$, leading to the existence of vortices carrying an arbitrary fraction of a flux quantum. It has also been argued that fractional flux quantum vortices will undergo a sharp Kosterlitz-Thouless transition, since for such vortices there is no exponential cut-off in the interaction between vortices (compare Abrikosov vortices, for which the exponential cut-off occurs for a separation distance larger than the effective penetration depth $\Lambda$). 

Comparing the width of the main peak in $\chi''(T)$ for film A, $\Delta T^{MgB_2}_{peak}\approx 0.05$ K with the corresponding result obtained for high quality  50 nm thick YBCO films, $\Delta T^{YBCO}_{peak}\approx 0.20$ K\cite{Festin1}, indicate that even in comparably thick MgB$_2$ films it is possible to observe, due to fractional flux quantum vortices, features reminiscent of a system undergoing a KT transition.

\begin{figure}
   \begin{center}
   \begin{tabular}{c}
   \includegraphics[height=8cm]{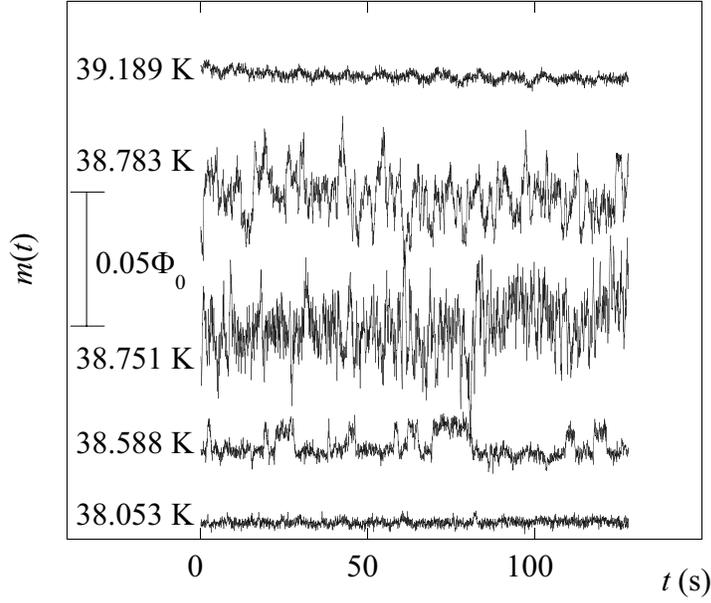}
   \end{tabular}
   \end{center}
   \caption[Figure 2.] 
   { \label{fig:fig2}Time trace $m(t)$ of the flux noise at five different temperatures for film A.
 } 
\end{figure}

Fig. 2 shows time traces $m(t)$ of the magnetic flux noise, i.e. the data used to calculate  $S_{\Phi}(f,T)$, at five different temperatures. The time trace corresponding to the highest temperature, 39.189 K, which is above the superconducting transition temperature, gives the background noise in the experimental setup. A weak sinusoidal noise contribution with a period of $\approx 5$ seconds can be observed in $m(t)$, originating from the resistive heater of the temperature control system in the experimental setup. $m(t)$ recorded at 38.053 K is obtained when the superconducting state in the film is fully developed. Again, the measured noise corresponds to the background noise of the experimental setup. $m(t)$ recorded at 38.783 K and 38.751 K correspond to flux noise measured close to the peak temperatures of $\chi''(T)$. Here, the correlated dynamics of vortex pairs is probed, yielding "coloured" flux noise spectra. The time trace measured at $T=38.588$ K exhibits features characteristic of ``random telegraph noise''. Comparing with  AC-susceptibility results, it can be seen that this noise corresponds to a comparably broad low temperature peak in $\chi''(T)$, which may indicate that the noise originates from Josephson vortices, formed between grains in the film, performing thermally activated jumps between two pinning sites\cite{Jung}.

Fig. 3 (a) shows flux noise spectra for film A at three different temperatures. For $T=38.751$ K, $f_{0A} \approx 1 $Hz and for $f > f_{0A}$, the spectrum in the investigated frequency range exhibits approximately a $f^{-1.4}$ dependence. Moreover, for $f<f_{0A}$, instead of the expected white noise, a weak frequency dependence is observed. This somewhat uncharacteristic behavior of the flux noise is connected with the multi-step transition observed for the MgB$_{2}$ films. This implies that for each temperature there are multiple macroscopic regions in the sample contributing to the measured noise and exhibiting different characteristic frequencies. At $T=38.783$ K, $f_{0A}$ from the lower temperature has increased by three orders of magnitude ($ \approx 10^{3}$ Hz). Moreover, at this temperature, a second and smaller characteristic frequency can clearly be resolved, $f_{0B} \approx 0.1 $Hz, which when increasing the temperature further to $T=38.751$ K increases in magnitude to $f_{0B} \approx 10^{2}$ Hz. It should be noted that $f_{0A}$ has increased above the measurement range at the highest temperature shown in this figure. The temperature evolution of $f_{0A}$ and $f_{0B}$ is more clearly seen in Fig. 3 (b), where $f \times S_{\Phi}$ vs. $f$ is shown. Moreover, by comparing to Fig. 1 (b), it can be seen that $f_{0A}$ and $f_{0B}$ correspond to the two peaks in $\chi''(T)$ being highest in temperature ($f_{0A}$ to the peak being largest in magnitude). 

\begin{figure}
   \begin{center}
   \begin{tabular}{c}
   \includegraphics[height=10cm]{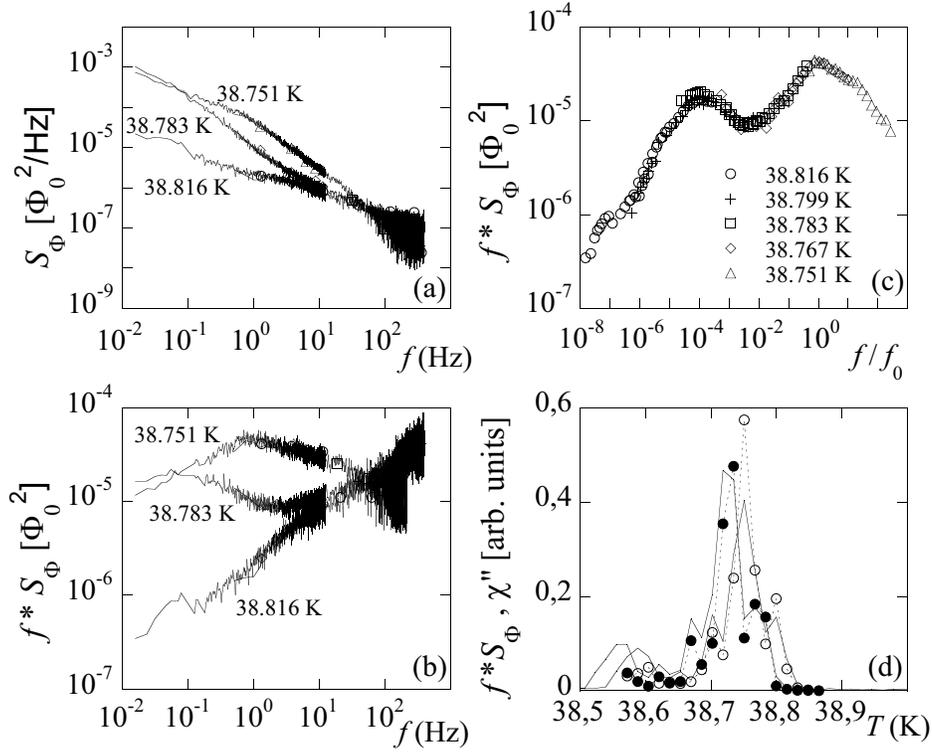}
   \end{tabular}
   \end{center}
   \caption[Figure 3.] 
   { \label{fig:fig3} 
Flux noise spectra recorded for film A. (a) $S_{\Phi}$ vs. $f$ at three different temperatures. (b) $f \times S_{\Phi}$ vs. $f$ for the same temperatures as in Fig. 3 (a). (c) Scaling plot, $f \times S_{\Phi}$ vs. $f/f_0$ including noise data obtained at five different temperatures. The extracted values of $f_0$, corresponding the $\chi''$ peak exhibiting the largest magnitude, are shown in Fig. 4 (a). (d) $f \times S_{\Phi}$ and $\chi''$ vs. temperature for two different frequencies, $f=17$ mHz and $1.7$ Hz. Solid lines are $\chi''$ data and dashed lines plus open and closed circles are $f \times S_{\Phi}$ data. } 
\end{figure} 

In a 2D superconductor, close to the KT transition, the following scaling relation is expected to hold; $f \times S_{\Phi}(f,T)=F(f/f_{0}(T))$, where $F$ is a scaling function\cite{Shaw}. In Fig. 3 (c), $f \times S_{\Phi}(f,T)$ is plotted as a function of $f/f_{0}(T)$ for five different temperatures. The  $f_{0}(T)$ values used in this scaling plot correspond to $f_{0A}(T)$ and are shown in Fig. 4 (a). The fact that $f_{0A}(T)$ can be used to scale {\em all} flux noise data, in spite of the complicated structure of the flux noise spectra with multiple characteristic frequencies at each temperature, strongly indicate that each "small" transition can be described with {\em the same} dynamics. This also implies that the different macroscopic regions in the film, having slightly different transition temperatures, are homogenous, i.e. that there are few pinning centers in the different regions affecting the random motion of vortices since they can all be scaled using the same characteristic frequencies $f_{0A}(T)$.

In Fig. 3 (d), $f \times S_{\Phi}(T)$ and $\chi''(T)$ are plotted as a function of temperature for two different frequencies, 17 mHz and 1.7 Hz. As can be seen in this figure, the flux noise data, which have been multiplied with a temperature and frequency independent factor, trace out the same characteristic features as do the AC-susceptibility data, confirming the validity of the fluctuation-dissipation-theorem.
 
In Fig. 4, $f_0(T)$ obtained from the peak in $\chi''(T)$ being largest in magnitude and from the two peaks closest in temperature on each side of the this main peak are plotted for both films. For film A, $f_{0}(T)$ obtained from flux noise data is also plotted. For each film, the temperature evolution of the different $f_0(T)$ corresponding to  different peaks in $\chi''(T)$ are within the experimental accuracy the same. For film A, $f_0$ increases from $17$ mHz to $51$ kHz in a temperature interval of only $\Delta T \approx 0.10$ K, while for film B $\Delta T \approx 0.18$ K. As a comparison $\Delta T$ for a high quality YBCO film\cite{Festin2} was found to be $\approx 0.2$ K. The difference in $f_{0}(T)$ for the two films can be explained by degradation of film B. Applying Eq. (1) to the $f_{0}(T)$ data, using $bz$ and $T_{KT}$ as fitting parameters, yields $bz\approx$4.3$\pm$0.3 for both films, which is in good agreement with previous results\cite{Shaw} obtained for Josephson junction arrays. This result gives further support for the presence of fractional flux quantum vortices and because of this the existence of a KT transition in these comparably thick MgB$_{2}$ films. For $f_0(T)$ corresponding to the largest peak in $\chi''(T)$, $T_{KT}$ for film A is $ \approx 0.2$ K below the peak temperature of $\chi''(170$Hz), while for film B, $T_{KT}$ is $\approx 0.35$ K below the temperature of the main peak of $\chi''(170$Hz).   

\begin{figure}
   \begin{center}
   \begin{tabular}{c}
   \includegraphics[height=6cm]{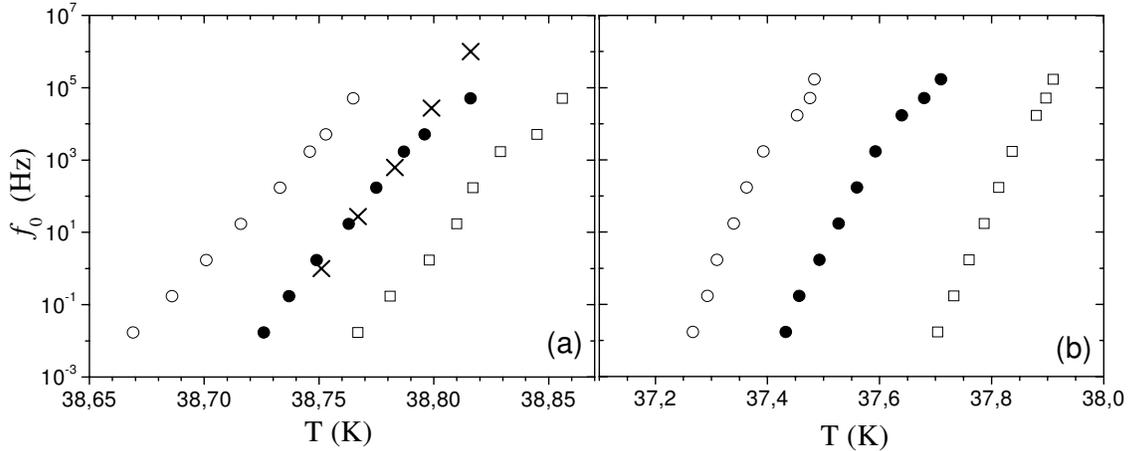}
   \end{tabular}
   \end{center}
   \caption[Figure 4.] 
   { \label{fig:fig4}$f_0$ vs. temperature obtained from the peak in $\chi''(T)$ being largest in magnitude (filled circles) and from the two peaks being closest in temperature on each side of the this main peak (open circles and squares). (a) Film A, crosses show the corresponding data obtained from the scaling plot shown in Fig. 3 (c). (b) Film B.} 
\end{figure} 

\section{Conclusions} 

We have performed flux noise and AC-susceptibility measurements on two 400nm thick MgB$_2$ films in the zero magnetic field limit. Both films show a superconducting transition having a multi step-like behavior in the real part of the AC-susceptibility with corresponding dissipation peaks in the imaginary component. In the flux noise data, as well as in the AC-susceptibility data, it is observed that there exists several characteristic frequencies, i.e. at each temperature there exists several characteristic  length scales describing the vortex correlations. This means that there are macroscopic regions in the films with slightly different critical temperatures. The width in temperature of each individual dissipation peak is less than half the width observed in high quality YBCO films,  therefore indicating a sharp $KT$ transition in the MgB$_{2}$ films. One plausible  explanation for observing features in the experimental results reminiscent of a KT transition is that there exists, due to two band superconductivity, fractional flux quantum vortices in the material. For ordinary Abrikosov vortices, the $KT$ transition would not be observed experimentally since there is an exponential cut-off in the logarithmic interaction for vortex separations being larger than the effective penetration depth $\Lambda$. For vortices carrying an arbitrary fraction of a flux quantum, no such cut-off exists, yielding a well-defined KT transition.

\section{Acknowledgements} This work was supported by the Swedish Research Council and the Ministry of Science and Technology of Korea through the creative Research Initiative Program. We are grateful to Egor Babaev for stimulating discussions relating to two band superconductivity and fractional flux quantum vortices.


\end{document}